\documentclass[fleqn,useAMS,usenatbib]{mnras}
\usepackage{amsmath,subfig,footmisc}
\usepackage{graphicx,txfonts,fleqn,enumerate,color,hyperref,listings}
\usepackage{ulem} 

\renewcommand*{\v}[1]  {\boldsymbol{#1}}

\newcommand*  {\Exp}[1]  {\mathrm{e}^{#1}}

\newcommand*  {\op}[1]   {{\hat{#1}}}

\newcommand   {\sub}[2]  {{#1}_{\mathrm{#2}}}

\newcommand*  {\twovector}[2] {{\begin{pmatrix} $1 \\ $2 \end{pmatrix}}}

\renewcommand {\emph}[1]  {\textit{#1}}

\title[{Multiple timestep integrators}]
{\boldmath Multiple timestep reversible $N$-body integrators for close encounters in planetary systems}

\author[David M. Hernandez and Walter Dehnen]
	{David M.~ Hernandez\href{http://orcid.org/0000-0001-7648-0926}{\includegraphics[width=11pt]{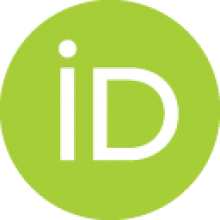}}$^{1}$\thanks{Email: david.m.hernandez@yale.edu} and 
    Walter Dehnen\href{http://orcid.org/0000-0001-8669-2316}{\includegraphics[width=11pt]{orcid-ID}}$^{2,3}$\\ 
	$^1$ Yale University, 52 Hillhouse, New Haven, CT 06511, USA \\
	$^2$ Astronomisches Rechen-Institut, Zentrum f\"ur Astronomie der Universit\"at Heidelberg, M\"onchhofstra\ss{}e 12-14, 69120, Heidelberg, Germany \\
    $^3$ School for Physics and Astronomy, University of Leicester, University Road, LE1 7RH, Leicester, UK
	}
\begin{document}

\maketitle

\label{first page}
\begin{abstract} 
We present new ``almost'' time-reversible integrators for solution of planetary systems consisting of ``planets'' and a dominant mass (``star'').  The algorithms can be considered adaptive generalizations of the Wisdom--Holman method, in which all pairs of planets can be assigned timesteps.  These timesteps, along with the global timestep, can be adapted time-reversibly, often at no appreciable additional compute cost, without sacrificing any of the long-term error benefits of the Wisdom--Holman method.  The method can also be considered a simpler and more flexible version of the \texttt{SYMBA} symplectic code.  We perform tests on several challenging problems with close encounters and find the reversible algorithms are up to $2.6$ times faster than a code based on \texttt{SYMBA}.  The codes presented here are available on Github.  We also find adapting a global timestep reversibly and discretely must be done in block-synchronized manner or similar.
\end{abstract}
\section{Introduction}
\label{sec:intro}
Orbital dynamics has been essential to the development of astronomy as a field.  Newton's Universal Gravitation \citep{Newton1687} provides a set of second order differential equations for the rate of change of gravitational systems.  Given initial conditions, if we are able to solve the set of equations, we can obtain the past or future (since the equations are time-symmetric) evolution of any gravitational system.  We have been able to effectuate this procedure to gain insight into the formation of galaxies, or understanding of the long-term stability of the Solar System; indeed, entire subfields of astronomy have been created as a result.

The mathematical problem we are describing is known as the $N$-body problem, in which $N$ point masses interact through pairwise forces: it has attracted substantial attention since Newton from scientists such as Laplace, Lagrange, and Poincar\'{e}.  The $N$-body problem has no closed-form solution in general.  One particular challenge of this problem is that solving it requires a multidisciplinary effort: first, we need to develop powerful astronomical tools capable of measuring the initial conditions needed, the theory of differential equations has to be developed to solve the system, chaotic dynamics must be understood to make sense of the calculations and the randomness involved, and finally, we need the development of technology (computers) which can rapidly carry out all the calculations.  

The requirement of all these ingredients has meant many of the most interesting astronomical problems have only been in reach in the last few decades.  One of the least understood parts of this process of solving the $N$-body problem is the role chaos plays in the accuracy of solutions \citep{Valtonen76,Smith77,Heggie91,QT92,PB14,BP15,Hernandezetal2020}.  We will not focus on this in this paper, but rather on the theory of the differential equations.

Dynamical astronomy made large progress due to the advent of symplectic methods \citep{DF76,Y90,chan90,kino91,WH91,sah92,Y93,cal93,WHT96,Chin97,preto99,mik99,mik99b,LR01,Y01,Y02,C02,ChinandChen05,hair06,rein11,Farresetal2013,Blanesetal2013,HB15,H16,Petitetal2019}.  Symplectic integrators solve Hamiltonian systems like $N$-body systems; they conserve exactly Poincar\'{e} invariants and are particularly well-suited to study systems undergoing many orbits over long timescales, and are thus used to probe the evolution of gravitational systems like galaxies or the Solar System.  Unfortunately, it is difficult to construct symplectic integrators that require changing timescales, such as in collisional systems with short relaxation times \citep{BT08}.  Conventional integration methods like the Runge--Kutta--Fehlberg method \citep{press02}, Bulisch--Stoer method \citep{press02}, or the Hermite integrator \citep{Makino1991} are far better suited for collisional systems with close encounters, but they lack the long-term error benefits of the symplectic methods.  Ideally we could combine the long-term advantages of symplectic methods with the adaptability of conventional methods.  Despite the tendency of symplectic methods to fail for collisional systems, with few good alternatives, they have remained popular.

Focusing in on the field of planetary dynamics, there have been several widely used, transformative symplectic methods.  The Wisdom--Holman method (WH)  \citep{WH91} is well-suited for nearly Keplerian systems like the Solar System.  \cite{C99,Reinetal2019} develop a hybrid symplectic integrator that switches between WH and another method when near-Keplerian motion ceases.  \cite{DLL98} (hereafter DLL98) offer another alternative, in which WH takes pairwise adaptive steps.  While not perfect, these methods have significantly advanced the study of planetary dynamics.

Time-reversible integrators are an alternative to symplectic integrators, and in many cases perform just as well in long-term simulations, while being as flexible as conventional methods \citep{HMM95,Funatoetal96,K98,HLR01,MP04,hair06,Makinoetal2006,HMS09,D17,HB18,Boekholtetal2022}.  As long as the method recovers the initial conditions when integrating forwards and then backwards, it is time-reversible.  Time-reversible methods have been largely ignored in practice, possibly due to their often large expense or the difficulty in implementing them.    It has recently been shown, however, that ``almost'' (which we refer to simple as time-reversible) time-reversible algorithms can switch between two methods, often at no penalty to compute time or accuracy \citep{HD2023}.

\cite{HD2023} used their simple time-reversible algorithm to devise a simpler, more flexible version of the symplectic hybrid integrator developed by \cite{C99,Reinetal2019} (with implementation known as \texttt{MERCURY}/\texttt{MERCURIUS}).  Currently, the \citeauthor{HD2023} algorithm is being implemented (Lu et al., in prep) as part of the \texttt{REBOUND} package \citep{RL12}.  Among its advantages over \texttt{MERCURY}/\texttt{MERCURIUS}, for example, it can successfully integrate close encounters with the dominant star mass. 

In this work, we similarly develop simple, fast, and flexible time-reversible versions of the DLL98 algorithm, implemented as \texttt{SYMBA}.  To accomplish this, we can generalize the algorithm of \cite{HD2023} to allow time-reversible switching between an arbitrary number of integrators (and thus timesteps).  The advantage of our approach is in the speed and simplicity of the method.  We develop two time-reversible methods, MTR (Multiple Timestep Reversible) and AG (Adaptive Global), and implement a \texttt{SYMBA}-like symplectic integrator, MTS (Multiple Timestep Symplectic), for fair comparisons.  In all our comparisons, the reversible algorithms are at least as accurate as MTS and one of the reversible methods is always faster.  This all holds while maintaining the benefits of more flexible adaptive conventional integrators.  For easy reference, the codes used in this work are provided at a Github link\footnote{\texttt{https://github.com/dmhernan/Reversible-Stepping} \label{Rstep}}. We warn the reader that they are not in a user friendly form and provided merely for reference.  We have tested the reversible methods on challenging problems involving close encounters throughout the paper.  Our codes can be considered as proof of concept, as they are not yet implemented in a faster compiled language like C; this is left for future work.

In Section \ref{sec:symp}, we provide background on developing integrators and on notation.  In Section \ref{sec:kep}, we develop both symplectic and time-reversible multiple timescale and timestep algorithms suitable for the two-body Kepler problem, with numerical demonstrations.  In Section \ref{sec:pcomp}, we generalize these methods to systems with multiple bodies with multiple timesteps varying independently.  We focus on systems with one dominant mass, the ``star.''  We conclude in Section \ref{sec:conc}.

\section{Background on symplectic and time-reversible integrators}
\label{sec:symp}
For this paper, we consider conservative Hamiltonian systems with form,
\begin{equation}
H = T(\v{p}) + V(\v{q}),
\label{eq:basicH}
\end{equation}
where $\v{p}$ and $\v{q}$ are the conjugate momenta and positions, respectively.  This form describes many orbital mechanics problems in astronomy.  More compactly, we describe these coordinates via $\v{z}(t) = (\v{p}(t),\v{q}(t) )$, with $t$ the time.  The equations governing them are,
\begin{equation}
\frac{d\v{z}}{dt} = \{\v{z}, H\},
\label{eq:solt}
\end{equation}
with formal solution,
\begin{equation}
\v{z}(t) = \exp(t \hat{H} ) \v{z}(0).
\label{eq:formals}
\end{equation}
$\{\}$ are Poisson brackets.  For functions $a(\v{z})$ and $b(\v{z})$, the following definition holds:
\begin{equation}
\{a,b\} = \sum_i \left( \frac{\partial a}{\partial q_i}\frac{\partial b}{\partial p_i} - \frac{\partial a}{\partial p_i}\frac{\partial b}{\partial q_i} \right).
\end{equation}  
The Lie operator of $H$ is $\hat{H}$ and is defined $\hat{H} a = \{a, H\} $.  Solution \eqref{eq:formals} is not often practical because it is not obtainable in closed-form.   Our task in this paper is to find more efficient and practical approximate solutions.  One approach is to take advantage of the function separation in eq. \eqref{eq:basicH} to obtain the ``map,''  
\begin{equation}
\v{z}_\mathrm{1}(t) = \exp(t \hat{T} ) \exp(t \hat{V} ) \v{z}(0) = \v{z}(t) + \mathcal O(t^2),
\label{eq:sol1}
\end{equation}
where $\v{z}_1$ indicates a first-order approximation in time\footnote{A small note on notation: in eq. \eqref{eq:sol1}, we really mean to indicate $ \exp(t \hat{T} ) \circ \exp(t \hat{V} )$: we evaluate $\v{z}^\prime = \exp(t \hat{V}) \v{z}(0)$, and the next Poisson brackets with $\exp(t \hat{T} )$ are taken with respect to $\v{z}^\prime$.  For simplicity, we omit ``$\circ$'' henceforth.  For even order methods like we consider in this paper, this technicality does not matter.  See also Appendix A in \cite{Tamayoetal2020}.}.  As an alternate map, switch $\hat{T}$ and $\hat{V}$\  

The solution of eq. \eqref{eq:sol1} is simple:
\begin{equation}
\label{eq:mapeul}
\v{q}_1(t) = \v{q}(0) + t \frac{\partial T}{\partial \v{p}}\bigg\rvert_{\v{p}_1} ~~~~ \mathrm{and} ~~~~\v{p}_1(t) = \v{p}(0) - t \frac{\partial V}{\partial \v{q}}\bigg\rvert_{\v{q}_0}.
\end{equation}
This map is symplectic because it preserves phase space volumes \citep{hair06} and, as a result, long-term orbits obtained with it are more reliable and accurate.  Also, map \eqref{eq:mapeul} is derived from differential equations themselves derived from a function $\tilde{H}_1$, which is just $H$ to $0$th order in $t$.  $\tilde{H}_1$ has form,
\begin{equation}
\tilde{H}_\mathrm{1} = H + \frac{t}{2} \{T,V\} + \frac{t^2}{12} \left( \{\{T,V\},V\} + \{\{V,T\},T\} \right) + \mathcal O (t^3).
\end{equation}
That the lowest power in $t$ is one indicates it is a first-order method.  Map \eqref{eq:sol1} is not time-reversible: integrating backwards does not recover the initial conditions.  In fact, any odd order method cannot be time-reversible.  Time-reversible methods are better than non-reversible methods at obtaining long-term accurate orbits \citep{hair06}.  For this work, we will not distinguish between time-symmetric and time-reversible methods.  A both symplectic and time-reversible map instead is shown as follows and known as leapfrog:
\begin{equation}
\v{z}_\mathrm{2}(h) = \exp\left(\frac{h}{2} \hat{T} \right) \exp(h \hat{V} ) \exp\left( \frac{h}{2} \hat{T} \right) \v{z}(t_0) = \v{z}(h) + \mathcal O(h^3).
\label{eq:sol2}
\end{equation}
Again, an alternate map switches $\hat{T}$ and $\hat{V}$:  
\begin{equation}
\v{z}^\prime_\mathrm{2}(h) = \exp\left(\frac{h}{2} \hat{V} \right) \exp(h \hat{T} ) \exp\left( \frac{h}{2} \hat{V} \right) \v{z}(t_0) = \hat{\zeta}(h) \v{z}(t_0).
\label{eq:sol2p}
\end{equation}
$h$ is now chosen as a small time interval $h = t-t_0$, the ``timestep,'' with $t_0$ the initial time.  The associated function $\tilde{H}_\mathrm{2}$ is now,
\begin{equation}
\tilde{H}_\mathrm{2} = H + \frac{h^2}{24} \left( -\{\{V,T\},T\} + 2 \{\{T,V\},V\} \right) + \mathcal O (h^4),
\label{eq:htild2}
\end{equation}
and $h$ must be small enough to ensure $\tilde{H}_2$ remains close to $H$.  A long-term solution is obtained by repeatedly applying map \eqref{eq:sol2}.

\subsection{Phase space dependent timesteps}
\label{sec:sympvar}
Leapfrog is no longer symplectic or time-reversible if we try to adapt the stepsize as a function of phase space $h = h(\v{z}(0) )$.  The symplecticity of map \eqref{eq:sol2} is defined by the condition,
\begin{equation}
\v{J}^{T} \v{\Omega} \v{J} = \v{\Omega}, 
\label{eq:scond}
\end{equation}
where $\v{J}$ is the Jacobian of $\v{z}(h)$ with respect to $\v{z}(0)$, and $\v{\Omega}$ is a constant anti-symmetric matrix whose form depends on the ordering within $\v{z}$.  When $h$ is no longer a constant, the condition \eqref{eq:scond} is generally destroyed.  That time-reversibility is lost is also seen immediately.  We show how to solve these problems in Section \ref{sec:MT}.

\section{Multiple timestep integrators for the Kepler problem}
\label{sec:kep}
\subsection{Symplectic algorithm}
\label{sec:MT}
\cite{SB94} showed a strategy to essentially adapt leapfrog's timestep in a symplectic way, by decomposing the interaction potential into a sum of components, each treated with a different timestep.  We describe the implementation of this idea by \cite{Leeetal1997}.  We consider a $2$D Kepler Hamiltonian.  Using Hamiltonian \eqref{eq:basicH}, choose the functions,
\begin{equation}
T = \frac{1}{2}\left(p_x^2 + p_y^2 \right),
\label{eq:T}
\end{equation}
and
\begin{equation}
V = -\frac{1}{{q}},
\label{eq:V}
\end{equation}
with $q = \sqrt{q_x^2 + q_y^2}$.  Then we select a set of cutoff radii $r_1 > r_2 > ...>r_n$ with constant ratio $r_i/r_{i + 1} = R$.  Each cutoff radius range, and by extension, potential range, has a corresponding timestep, such that if $q > r_1$, $h = h_0$; if $r_1 > q > r_2$, $h = h_1$; and so on until $h = h_{n - 1}$.  Also, $h_i/h_{i + 1} = M$ and the global timestep is $h_0$.  Now we decompose the potential into,
\begin{equation}
\label{eq:Vdec}
V = \sum_{i = 0}^\infty V_i = \tilde{V}_0 + (\tilde{V}_1 - \tilde{V}_0 ) + (\tilde{V}_2 - \tilde{V}_1) + \hdots,
\end{equation}
where $V_0 = \tilde{V}_0$, $V_1 = \tilde{V}_1 - \tilde{V}_0$, $V_2 = \tilde{V}_2 - \tilde{V}_1$, $\hdots$, with associated forces, $\tilde{\v{F}}_i =  -{\partial \tilde{V}_i}/ {\partial \v{q}}$.  When an operator $\hat{V}_i$ appears, we substitute in terms of $\hat{\tilde{V}}_j$.  We focus on the form of these forces adopted in the tests of DLL98:
\begin{equation}
    \tilde{\v{F}}_{i - 1}= 
\begin{cases}
    -\frac{\v{q}}{q^3}& \text{if } q\geq r_i,\\
    -f\left(\frac{r_i - q}{r_i - r_{i + 1}}\right)\frac{\v{q}}{q^3}              & \text{if } r_{i + 1} \leq q < r_i, \text{ and} \\
    0 & \text{if } q< r_{i+1},
\end{cases}
\label{eq:ftild}
\end{equation}
where $f(x) = 2x^3 - 3x^2 + 1$.  $\tilde{\v{F}}_i$ transitions smoothly from $0$ to $-\v{q}/q^3$.  This polynomial has the properties $f(1) = \frac{df}{dx}\big\rvert_{0} =  \frac{df}{dx}\big\rvert_{1} = 0$ and $f(0) = 1$.  This form ensures $V(q)$ has continuous derivatives, and thus the symplecticity of methods containing operators $\hat{V}_i$ is ensured \citep{H19}.  Fig. 1 of DLL98 and Fig. 3 of \cite{Leeetal1997} plot the force decompositions $F_i$ for forces in a single dimension.

Now, define,
\begin{equation}
\hat{\phi}_i = \exp\left(h_i \left(\sum_{j = i}^\infty \hat{V}_j + \hat{T} \right) \right).
\label{eq:phi}
\end{equation}
We approximate the $\hat{\phi}_i$ using leapfrog methods $\hat{\tilde{\phi}}_i$, such that $\hat{\tilde{\phi}}_i =  \hat{{\phi}}_i + \mathcal O(h^2)$, but the two sets of operators in eq. \eqref{eq:sol2} are now $\hat{\tilde{\phi}}_{i+1}$ and $\hat{V_i}$ or $\hat{V}_{i_m}$ and $\hat{T}$:
\begin{equation}
\begin{aligned}
\hat{\tilde{\phi}}_{i_m} &=  \left[\exp\left(\frac{h_{i_m}}{2} \hat{V}_{i_m}\right) \exp\left( h_{i_m} \hat{T}\right) \exp\left(\frac{h_{i_m}}{2} \hat{V}_{i_m}\right) \right]^M,\label{1}\\
\hat{\tilde{\phi}}_i &= \left[\exp\left(\frac{h_i}{2} \hat{V}_i\right) \hat{\tilde{\phi}}_{i+1} \exp\left(\frac{h_i}{2} \hat{V}_i\right) \right]^M\text{  for $0 < i < i_m$, and}\label{2}\\
\hat{\tilde{\phi}}_0 &= \exp\left(\frac{h_0}{2} \hat{V}_0\right) \hat{\tilde{\phi}}_1 \exp\left(\frac{h_0}{2} \hat{V}_0\right). \label{3}
\end{aligned}
\label{eq:multt}
\end{equation}
$\hat{\tilde{\phi}}_0$ is the \cite{Leeetal1997} map.  $i_m$ is a maximum recursion level which may vary during evaluation of $\hat{\tilde{\phi}}_0$.  More details are provided in Section \ref{sec:imp}.  Note each $\hat{\tilde{\phi}}_i$ is $M$ leapfrogs with stepsize $h_i = h_0/M^i$.  In principle, the recursion of map \eqref{eq:multt} is infinite, but we can truncate it at the ${i_m}$ if we can be sure $q > r_{i_m+1}$ during global step $h_0$.  At best, we can only estimate the maximum $h_i$ during a global step; we outline such an estimate in Section \ref{sec:imp}.  That we can only estimate the maximum $h_i$ is a disadvantage of this method that will be avoided with the reversible scheme derived in Section \ref{sec:tvar}.  Map \eqref{eq:multt} allows us to take adaptive stepsizes while maintaining symplecticity and forms the basis of the \texttt{SYMBA} code (DLL98).  We have independently implemented map \eqref{eq:multt} and name it MTS for Multiple Timestep Symplectic.  

DLL98 derive a function $\tilde{H}$ from which the differential equations of the map can be derived, and is close to $H$:  
\begin{equation}
\begin{aligned}
\tilde{H} &= H + \sum_{i = 0}^\infty \frac{h_i^2}{12} \left\{ \left\{V_i, T\right\}, T + \frac{1}{2} V_i + \sum_{j = i+1}^\infty V_j \right\} + \mathcal O(h_i^4) \\
&= H + H_2 + \mathcal O(h_i^4) \\
&= H + H_{\mathrm{err}}^{\mathrm{Lee}}.
\end{aligned}
\label{eq:err}
\end{equation}
$H_{\mathrm{err}}^{\mathrm{Lee}}$ constitutes the error Hamiltonian, which must be kept small.  Define $\v{F}_{i} = -{\partial {V}_{i}}/ {\partial \v{q}} =  \tilde{\v{F}}_{i}  - \tilde{\v{F}}_{i-1}$.  Assume at some point in time $q \approx r_{i+1}$.  Then, from eq. \eqref{eq:ftild}, $\v{F}_{i} = -\v{q}/q^3$, while $\v{F}_{i-1} = \v{F}_{i+1} = \v{0}$.  It follows that,
\begin{equation}
H_2 = \frac{1}{12} \left(  h_i^2    \left\{ \left\{V_i, T\right\}, T + \frac{1}{2} V_i \right\} \right).                                                              
\label{eq:Herrsimp}
\end{equation}
We have $q \propto p^{-1/2}$, which implies, from eq. \eqref{eq:Herrsimp}, that $H_{\mathrm{err}}^{\mathrm{Lee}} = \mathcal O(h_i^2 q^{-4})$.  Thus, choosing $R = \sqrt{M}$ makes $H_2$ independent of the close encounter strength and $H_{\mathrm{err}}^{\mathrm{Lee}} = \mathcal O(h_0^2)$.  Another logical choice is to set $M=3$ and $R=2$, which yields $h \propto q^{3/2}$, a timestep proportional to the free-fall time, so that the phase is resolved evenly during free-fall.  This gives $H_{\mathrm{err}}^{\mathrm{Lee}} = \mathcal O(h_0^2 q^{-1})$, adequate if the encounter is not extremely close.  In addition, the smoothness of $f$ ensures the $h_i^2$ term is defined.  Making $f$ even smoother (in the sense that higher order derivatives of $f$ are $0$ at $x=0$ and $x=1$) can increase the accuracy of the map \citep{H19b}, and increases the number of terms at higher powers of $h_i$ well-defined in $\tilde{H}$.

\subsubsection{Estimating the maximum recursion level}
\label{sec:imp}
The code \texttt{SYMBA} estimates the maximum $h_i$ required during a global step \eqref{eq:multt}, as explained above.  We describe this implementation here, and also implement it in MTS.  For full details, refer to our code\footref{Rstep}.  If any of the following conditions are satisfied before evaluating $\hat{\tilde{\phi}}_{i}$, solve this operator and require recursion at $\hat{\tilde{\phi}}_{i+1}$.  Otherwise, set $i = i_m$.   

\begin{itemize}

\item 
If $q < r_{i+1}$.

\item
If $\v{q} \cdot \v{p} < 0$, compute an estimate of the time to shortest separation, $t_{\mathrm{min}} = -\v{q} \cdot \v{p}/p^2$.  If $t_{\mathrm{min}} < h_i$, let $r_{\mathrm{min}} = \sqrt{q^2 - (\v{q} \cdot \v{p})^2/p^2  } $.  ``If $ \mathrm{min} (r_{\mathrm{min}}, q) < r_{i}$'' is the condition.  If $t_{\mathrm{min}} > h_i$, let $r_{\mathrm{min}} = \sqrt{q^2 + 2 h_i \v{p}   \cdot \v{q} + p^2 + h_i^2} $.  ``If $\mathrm{min} (r_{\mathrm{min}}, q) < r_{i}$'' is the condition.

\end{itemize}
Our estimates only use positions and velocities/momenta, and no higher time derivatives, but work well enough in practice.  One inefficiency of our implementation MTS is that once we have entered a recursion map $\hat{\tilde{\phi}}_i$, the maximum recursion level for the remainder of the global timestep $h_0$ must be at least $i$ and cannot decrease.  A more efficient algorithm would allow $i$ to decrease.  Our goal is to implement a simple and accurate MTS method, and the focus of this paper is on reversible, not symplectic, methods.

\subsection{Time-reversible algorithms}
\label{sec:tvar}
\subsubsection{Adapting pairwise timesteps}
We now describe an algorithm which is the first novelty of this paper.  We wish to have a simpler multiple timestepping algorithm which avoids smoothing functions and can adapt the time-step not only based on the position.  The algorithm will be time-reversible but non-symplectic, which is just as good in many cases.  We call it MTR for Multiple Timestep Reversible.  This algorithm is a generalization of reversible maps from \cite{HD2023} which only switch between two global timesteps.  MTS will handle multiple adaptive pairwise steps.  First, we substitute the timestep criteria $r_i$ for timestep criteria $g_i$.  $g(\v{q},\v{p})$ is a useful possibly momentum-dependent function such as a free-fall time or a semi-major axis.  The timestep level $i$ is then determined by the condition $g_{i+1} < g(\v{q},\v{p}) < g_i$. As long as we set $g_0$, the $g_i$ follow from the adopted $R$ value.  Simply setting $g_i = r_i$ works for a variety of problems and, indeed, in this paper we assume this choice unless otherwise stated.  Define the map,
\begin{equation}
\hat{C}_i = \left[\exp\left(\frac{h_i}{2} \hat{V}\right) \exp\left({h_i} \hat{T}\right) \exp\left(\frac{h_i}{2} \hat{V}\right)\right]^{M^i},
\label{eq:Ckep}
\end{equation}
which is composed of $M^i$ substeps and evolves for global step $h_0$.  A timestep level $j_k$ is collected after each substep $k$.  Then, the condition for using $\hat{C}_i$ and $h_i$ is $i = \mathrm{max}_k \{j_k\}$.  For example, consider $\hat{C}_1$, with $M=3$,
\begin{equation}
\begin{aligned}
\hat{C}_1 &= \left[\exp\left(\frac{h_1}{2} \hat{V}\right) \exp\left({h_1} \hat{T}\right) \exp\left(\frac{h_1}{2} \hat{V}\right)\right] \\
&\times\left[\exp\left(\frac{h_1}{2} \hat{V}\right) \exp\left({h_1} \hat{T}\right) \exp\left(\frac{h_1}{2} \hat{V}\right)\right] \\
&\times \left[\exp\left(\frac{h_1}{2} \hat{V}\right) \exp\left({h_1} \hat{T}\right) \exp\left(\frac{h_1}{2} \hat{V}\right)\right].
\end{aligned}
\end{equation}
We collect timestep levels at three points.  For $\hat{C}_2$, we would collect nine timestep levels.  On occasion, global timesteps must be repeated once or even more when one of the $h_j$ collected at the substeps is less than $h_i$.  An algorithm demonstrating MTR is Listing \ref{lst:adapglobal}.  The \texttt{while} loop rarely requires more than one pass; it makes sure the maximum level has not changed after redoing a step.  The initial timestep level at time $0$ is calculated from the initial conditions.  Every time MTR is deciding between two $\hat{C}_i$ maps, it is subject to rare reversibility errors described in \cite{HD2023}; the so called ``ambiguous'' and ``inconsistent'' steps.

Listing \ref{lst:adapglobalp} shows an alternate implementation of MTR, which gives identical results.  We denote it \texttt{MTRa}.  \texttt{zeta} is given by eq. \eqref{eq:sol2p} and is a function of the state $\v{z}_1$ and substep $h_0/M^i$.  The \texttt{level} function calculates the timestep level.  We present {MTRa} to aid in understanding MTR.
\begin{lstlisting}[label={lst:adapglobal},caption={
\texttt{Python} script for the MTR time-reversible algorithm, the first main result of this paper.  MTR redoes timesteps as needed to enforce time-reversibility.
},
basicstyle=\ttfamily\footnotesize,
breaklines,
language=Python]
01 def MTR(z0,h0,i):
02 #integrate using h0,i giving z2(t=t+h0)
03     z1,levels=mapC(z0,h0,i)
04 #using macpC(), compute levels[k] = j 
05 #satisfying g_{j+1} < g(z(t_k)) < g_j for
06 #times t_k (every substep up to M^i substep)
07     iN = max(levels)
08     while iN > i
09         i = iN
10         z1,levels=mapC(z0,h0,i)
11         iN = max(levels)
12 # levels[-1] is used as input in next timestep
13     return z2,levels[-1]
\end{lstlisting}
\begin{lstlisting}[label={lst:adapglobalp},caption={
\texttt{Python} script for MTRa, a different implementation of MTR.
},
basicstyle=\ttfamily\footnotesize,
breaklines,
language=Python]
01  def MTRa(z0,i0,h0,M):
02      while True:
03          num_steps = M**i0
04          z1 = z0
05          for n in range(num_steps):
06              z1 = zeta(z1,h0/num_steps)
07              i1 = level(z1)
08              if i1 > i0:
09                  break
10          if i1 <= i0:
11              return z1,i1
12          i0 = i1
\end{lstlisting}

\subsubsection{Adapting the global step itself}
\label{sec:adapglob}

So far, we have kept $h_0$ fixed, but now we show how to adapt it reversibly.  In a system with a hierarchy of timesteps (see Section \ref{sec:kep}), perhaps we could set $h_0$ as the lowest timestep being used by any pair.  In a first example, we vary $h_0$ and do not use any other timestep, which is sufficient for the Kepler problem.  We need only consider the leapfrog method, eq. \eqref{eq:sol2}, $\hat{\phi}_0$, which we write as \texttt{mapphi0} in our algorithm.  We still increase or decrease $h_0$ via a factor $M$.  $h_0$ is easily reduced now when necessary, but we have found that increasing $h_0$ must be done with care.  $h_0$ is only increased if the number of $h_0$ steps taken modulo $M$ is $0$.  This algorithm does not exactly change steps at block synchronized points.  We have found this requirement via experimentation.  \cite{D17}, Section $6.1.1$, makes brief remarks about a similar finding, and provides a possible explanation.  Listing \ref{lst:adapth0} shows explicitly how this algorithm works and is our second main result, labeled AG for Adaptive Global.  \texttt{lev} is an array of the number of steps taken each timestep level, modulo $M$.
\begin{minipage}{\linewidth}
\begin{lstlisting}[label={lst:adapth0},caption={
\texttt{Python} script for the AG time-reversible algorithm, the second main result of this paper.  In contrast to MTR, AG varies the global timestep reversibly, and does not allow an increase of the stepsize at every point in time, which is required to maintain long-term accuracy. 
},
basicstyle=\ttfamily\footnotesize,
breaklines,
language=Python]
01 def AG(z0,h,i,M,lev):
02 #h is a vector of possible timesteps
03 #lev is a 0 vector at t=0
04 #M is the number of substeps per level
05 #Integrate using h[i], giving z2(t=t+h[i]) 
06 #j satisfies g_{j+1} < g(z1) < g_j.  
07     z1 = zeta(z0,h[i])     
08     j = level(z1)
09     if (i < j):
10         z1 = zeta(z0,h[j])
11         lev[j] += 1
12         return z2,j,lev 
13     else:
14         lev[i] += 1
15         while(i > j):
16 # increase step depending on modulus operator
17             i -= 1 if lev[i]%M == 0 else break  
18         end
19         return z2,i,lev
\end{lstlisting}
\end{minipage}
Listing \ref{lst:adapth0p} shows an alternate algorithm in which now the steps can only change at block synchronized points.  Although we do not present results from it in this paper, our tests indicate it is just as accurate as the algorithm of Listing \ref{lst:adapth0}.  Moreover, Listing \ref{lst:adapth0p} lends itself easily to be combined with Listing \ref{lst:adapglobal}, as shown by preliminary tests (not shown in this work).
\begin{minipage}{\linewidth}
\begin{lstlisting}[label={lst:adapth0p},caption={
\texttt{Python} script for the blockstep method, which varies the (global) time step at block synchronized points, and also maintains long-term accuracy.},
basicstyle=\ttfamily\footnotesize,breaklines,language=Python]
01  def block_step(z0,i0,hi,i,M):
02      if i0 <= i:   # initial level okay: attempt hi
03          z1 = zeta(z0,hi)
04          i1 = level(z1)
05          if i1 <= i:
06              return z1,i1
07          i0 = i1   # attempt failed: need higher level
08      z1,i1 = z0,i0
09      for m in range(M):
10          z1,i1 = block_step(z1,i1,hi/M,i+1,M)
11      return z1,i1
12  def AGa(z0,i0,h0,M):
13      return block_step(z0,i0,h0,0,M)
\end{lstlisting}
\end{minipage}

\subsection{Numerical comparisons among algorithms}
\label{sec:kepcomp}
\subsubsection{$0.9$ eccentricity}
\label{sec:e9}

\begin{figure}
	\includegraphics[width=90mm]{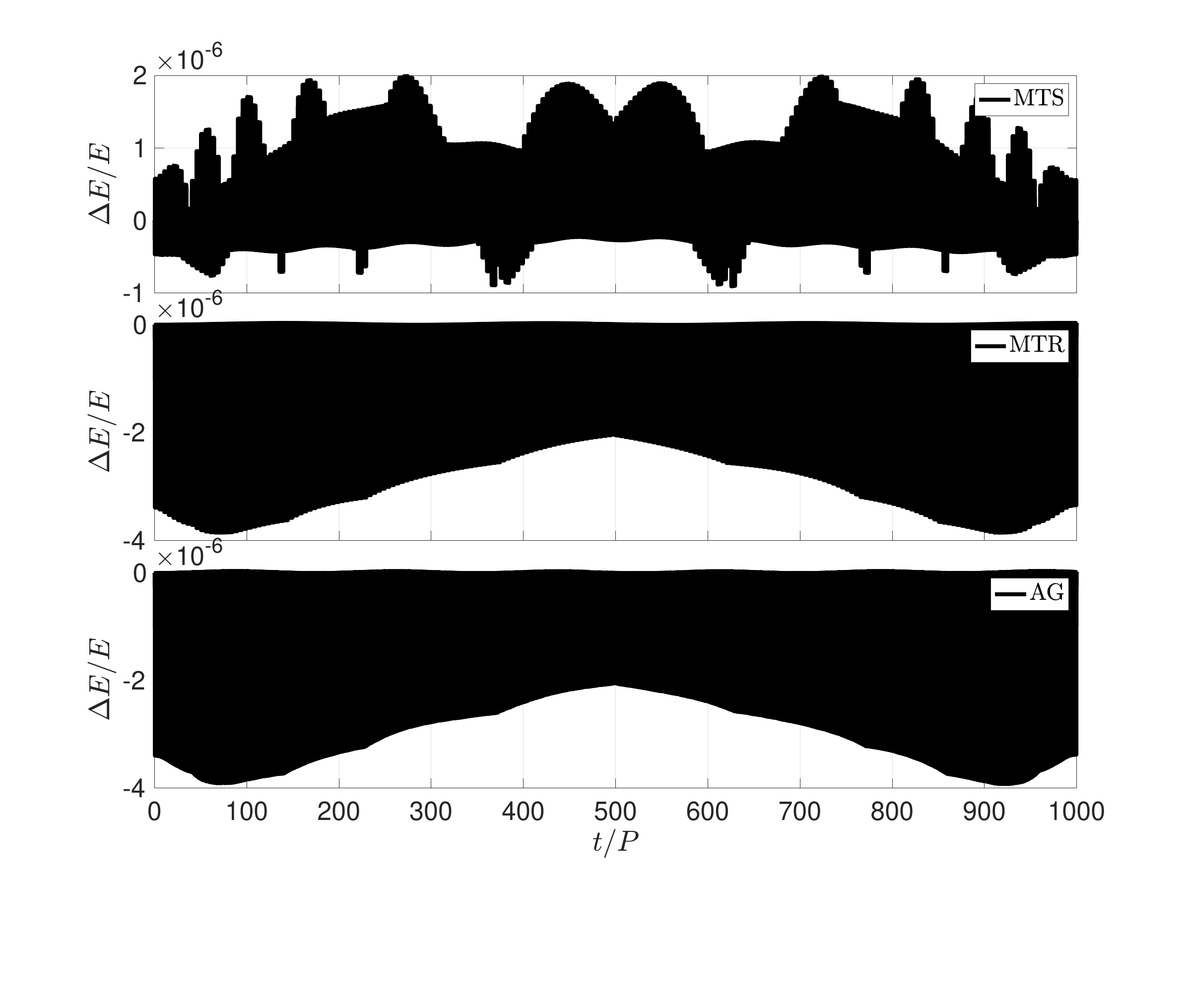}
	
	\caption{A first test of the near explicit time-reversible methods MTR and AG.  A Kepler orbit with $e = 0.9$ is integrated for $1000$ periods $P$, and the energy error is recorded over time.  We compare results to the \texttt{SYMBA}-like MTS method.  The errors are all stable over time and the reversible methods are the fastest algorithms.
	\label{fig:kepcomp}
  	}
\end{figure}

We now compare the performance of the symplectic MTS method and reversible MTR and AG methods.  It is most advantageous to use MTR for systems with more than two planets, as presented in Section \ref{sec:MRmany} , but we test it here for completeness.  We consider the 2D Kepler problem, equations \eqref{eq:basicH}, \eqref{eq:T}, and \eqref{eq:V}.  We set $R = \sqrt{2}$ and $M=2$ a suggested in Section \ref{sec:MT}, which is a steep scaling.  Also, we set the semi-major axis as $a=1$, $r_1 = \sqrt{2}$, the eccentricity as $e = 0.9$, and $h_0 = P/20000$, with $P$ the period ($= 2 \pi$).  $R$, $M$, $e$, and $h_0$ are chosen to match the parameters of \cite{Leeetal1997}, Fig. 4.  The initial conditions are at apocenter.  The energy error as a function of time is plotted in Fig. \ref{fig:kepcomp}.  Here $E$ is the energy (eq. \eqref{eq:basicH}) so $\Delta E/E$ is the relative energy error.  The evolution is calculated up to time $t = 1000P$ at $10 000$ linearly spaced outputs. 

The algorithms are implemented in the MATLAB language and the compute times for MTS, MTR, and AG are $23$, $13$, and $9$ s, respectively.  The relative compute times are a useful guide for the result in faster compiled languages.  Of course, these numbers depend on the processor being used but the important quantity to note are relative compute times, which are a fair comparison.  The reversible methods have a speed advantage.

All methods keep a small energy error over time which is not drifting secularly.  The errors of MTR and AG are not identical, although they appear close.  The tendency of the error to be more positive or negative is a function of the phase of the orbit of the initial conditions, which we verified.  MTR redid a fraction $4\times 10^{-4}$ of the timesteps ($800$ out of $2000001$ steps).  The fraction for AG was $6 \times 10^{-3}$ ($8000$ out of $13309460$ steps).  The smallest timestep used by any algorithm was $h_9 = P/1024000$, near pericenters.  For MTR, the timestep level between adjacent global timestep jumps by $\pm 1$ always.  The global timestep level for AG also jumps by $\pm 1$ always.

\subsubsection{$0.999$ eccentricity}
\label{sec:e999}
We next want to test extreme eccentricity orbits, which will require extremely small steps near pericenter.  The motivation for this extreme orbit is that \cite{Leeetal1997} also test extreme eccentricity orbits.  Except for setting $e = 0.999$, we use the same parameters and algorithms as in Section \ref{sec:e9}.  This test will require timestep level changes by more than one for MTR.  This happens because timestep levels near pericenter can vary significantly during a global timestep.  The MTS maximum recursion algorithm of Section \ref{sec:imp} becomes important for efficiency and catching rapid timestep level changes.  We expect the efficiency of MTR to be worse for $e = 0.999$ compared to $e=0.9$ since the timestep levels cannot be varied within a global step.  The runtimes are $2.7 \times 10^3$, $170$, and $270$ s for MTR, AG, and MTS, respectively, so the runtime of AG is $60\%$ that of MTS, while MTR is about $10$ times slower than MTS.  The fastest method is a reversible method.  

\begin{figure}
	\includegraphics[width=90mm]{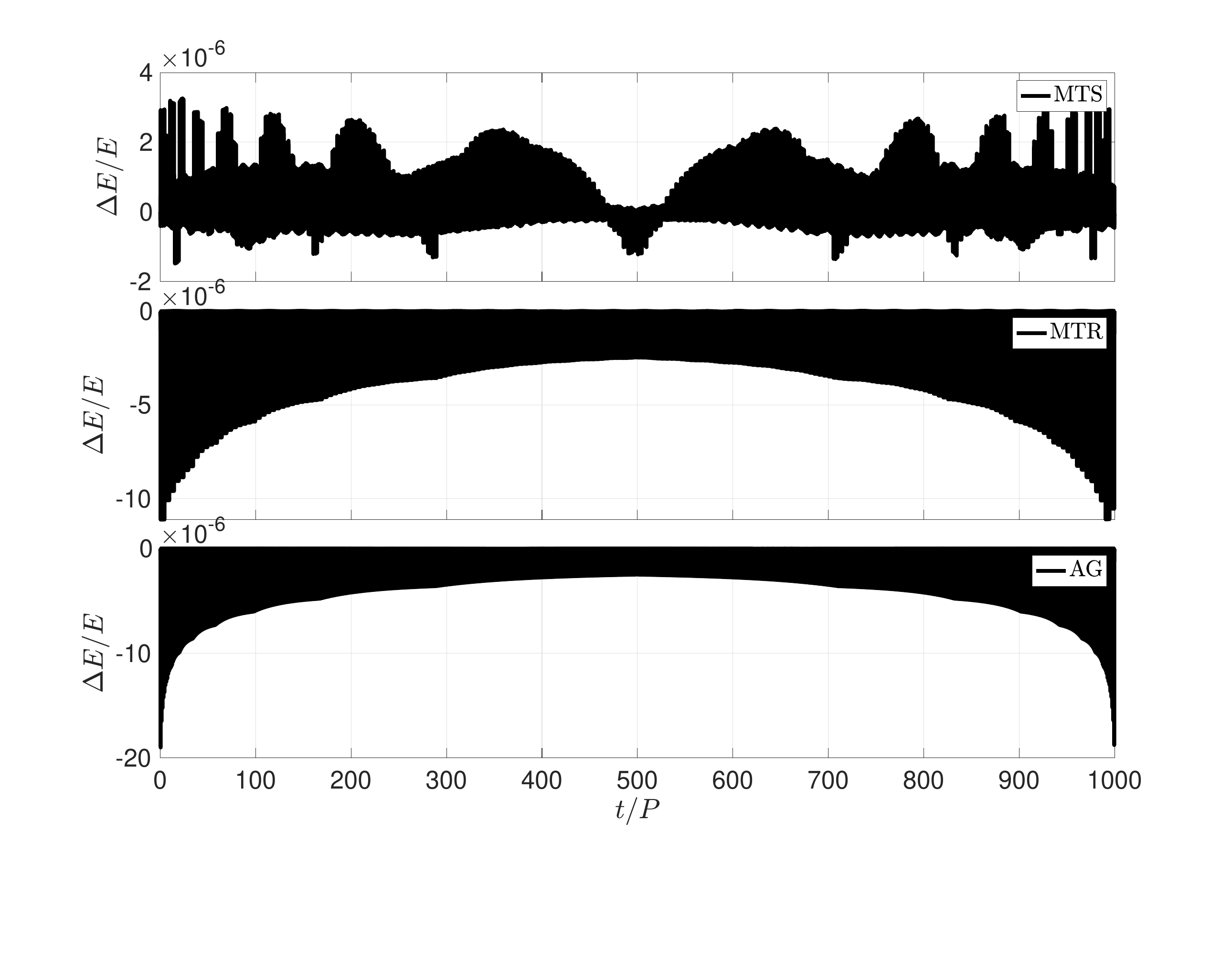}
	
	\caption{
	The same test as in Fig. \ref{fig:kepcomp}, except now with $e = 0.999$.  The errors over time for all methods are again comparable, and the fastest method is AG.
	\label{fig:energy999}
  	}
\end{figure}

We plot the energy error over time for the methods in Fig. \ref{fig:energy999}.
The errors over time for all methods again remain small.  In this case, the error amplitudes for AG and MTR are slightly larger in magnitude than those of MTS.  The median errors are $-5.5 \times 10^{-8}$, $-2.0 \times 10^{-7}$, and $2.0 \times 10^{-7}$ for MTS, MTR, and AG, respectively: the reversible methods have absolute median errors $3.7$ times that of MTS.  The smallest timestep used is $h_0/2^{22}$ (the $22$nd timestep level).  For MTR, timestep level changes by up to $\pm 10$ between global timesteps occur.

\section{Multiple timestep algorithms for planetary systems with a dominant mass}
\label{sec:pcomp}
In Section \ref{sec:kep} we showed that we can reversibly adapt the timestep in a single orbit with two different approaches, MTR and AG.  AG was faster than a symplectic \texttt{SYMBA}-like algorithm with adaptive steps.  In this section, we generalize the algorithms to work for systems with multiple bodies which have a dominant mass which we call ``Sun.''  The other bodies are called ``planets.''  For simplicity, we allow close encounters between planets only, and we don't allow close Sun--planet encounters.  For problems with close encounters with the Sun, solutions are presented elsewhere \citep[][Lu et al., in prep]{LD00}.  Each pair of planets will be evolved on different timescales, so that we can think of each pair of planets having a different timestep.  We begin with our 3D Hamiltonian of form \eqref{eq:basicH}:
\begin{equation}
H = T + V = \sum_i \frac{p_i^2}{2 m_i} - \sum_{i  < j} \frac{G m_i m_j}{q_{i j}}.
\label{eq:nbodyCart}
\end{equation}
Here, $m_i$ are masses, $G$ is the gravitational constant, and $q_{ij} = |\v{q}_i - \v{q}_j |$.  $(\v{q},\v{p})$ are the inertial Cartesian canonical coordinates.  Although we could proceed using Hamiltonian \eqref{eq:nbodyCart}, it has the disadvantage that $N$-body maps based on it perform poorly in frames that are not at the center of mass.  Thus, we recast Hamiltonian \eqref{eq:nbodyCart} in Democratic Heliocentric Coordinates (DLL98), denoted by $(\v{Q},\v{P})$.  $\v{Q}_0$ is the position of the center of mass and $\v{P}_0$ is the total momentum.  For $i > 0$, $\v{Q}_i$ are heliocentric positions and $\v{P}_i$ are baryentric momenta.  In these coordinates, $\v{Q}_0$ is cyclic and we can ignore a bulk kinetic term $P_0^2/(2M)$ in the Hamiltonian, with $M = \sum_i m_i$.  The total Hamiltonian is $H = H_{\mathrm{Kep}} + H_{\mathrm{Sun}} + V$, where
\begin{equation}
\begin{aligned}
H_{\mathrm{Kep}} =& \sum_{i \ne 0} H_{\mathrm{Kep},i} = \sum_{i \ne 0} \left(\frac{P_i^2}{2 m_i} - \frac{G m_i m_j}{Q_i}  \right), \\
H_{\mathrm{Sun}} =& \frac{1}{2 m_0} \left( \sum_{i \ne 0} P_i \right)^2, \text{and} \\
V =& - \sum_{0 < i < j} \frac{G m_i m_j}{Q_{i j}}.
\end{aligned}
\end{equation}
Here, $Q_{ij} = |\v{Q}_i - \v{Q}_j|$.  $H_{\mathrm{Sun}}$ is the Solar kinetic energy.  We are now ready to construct a second order map that solves this Hamiltonian, known as the Wisdom--Holman map \citep{WH91}.  In analogy to eq. \eqref{eq:sol2} for leapfrog, we have, 
\begin{equation}
z(h) = \exp \left(\frac{h}{2}  \hat{H}_{\mathrm{Sun}} \right) \exp\left(\frac{h}{2}  \hat{V}_{} \right) \exp\left(h \hat{H}_{\mathrm{Kep}} \right)  \exp\left(\frac{h}{2}  \hat{V}_{} \right)  \exp\left(\frac{h}{2} \hat{H}_{\mathrm{Sun}}\right)  z(0).
\label{eq:psi2}
\end{equation}
Because $\{H_{\mathrm{Sun}}, V\} = 0 $, the order we apply $\hat{H}_{\mathrm{Sun}}$ and $\hat{V}$ in is irrelevant.  The Wisdom--Holman map \eqref{eq:psi2} is efficient for studying planetary systems in which the planetary orbits remain well separated and thus $|H_{\mathrm{Kep}}| \gg |V| $ and $|H_{\mathrm{Kep}}| \gg |H_{\mathrm{Sun}}|$.  Solving $H_{\mathrm{Kep}}$ involves employing an incremental implicit Kepler equation solver.  We use the approach in Universal Variables, such that the equations of motion no longer have singularities at $Q_i = 0$.  Our implementation is described in \cite{WH15}.

\subsection{Symplectic integrator (DLL98)}
\label{sec:sympmany}
We now allow all the pairs of planetary bodies in eq. \eqref{eq:psi2} to be evolved on different timescales.  While this was first described in DLL98, here we provide more details and fill in gaps in the derivation.  Each pairwise distance will now be classified into a radius range (and thus, timestep level), which transitions smoothly into other shells using eq. \eqref{eq:ftild}.  First, rewrite $V$ in analogy to eq. \eqref{eq:Vdec} as, 
\begin{equation}
V = \sum_{k} V_k = \sum_{0< i<j} \tilde{V}_{ij} = \sum_{k,0 < i < j} V_{k i j}.
\label{eq:Vk}
\end{equation}
Here, $k$ is a shell index and $(i,j)$ labels the planetary pair.  Now, we update map \eqref{eq:multt} using the fact that,  
\begin{equation}
[\hat{V}_{ijk},\hat{V}_{lmn}] = 0,
\label{eq:commute}
\end{equation}
for any indices $i,j,k,l,m,n$: 
\begin{equation}
\begin{aligned}
\hat{\tilde{\psi}}_{k_m} &= \left[ \exp\left(\frac{h_{k_m}}{2} \sum_{(i,j) \text{ pairs}}\hat{V}_{k_{m}ij}\right)  \exp\left(h_{k_m} \sum_{i \in \v{W}_{k_m}} \hat{H}_{\mathrm{Kep},i} \right) \exp\left(\frac{h_{k_m}}{2} \sum_{(i,j) \text{ pairs}}\hat{V}_{k_{m}ij}\right)\right]^M,\\
\hat{\tilde{\psi}}_k &= \left[\exp\left(\frac{h_{k}}{2} \sum_{(i,j) \text{ pairs}}\hat{V}_{kij}\right) \exp\left(h_{k} \sum_{i \in \v{W}_{k}} \hat{H}_{\mathrm{Kep},i} \right) \hat{\tilde{\psi}}_{k+1} \right. \\
& \times \left. \exp\left(\frac{h_{k}}{2} \sum_{(i,j) \text{ pairs}}\hat{V}_{kij}\right)\right]^M\text{  for $0< k < k_m$, and}\\
\hat{\tilde{\psi}}_0 &= \exp\left(\frac{h_0}{2} \hat{H}_{\mathrm{Sun}}\right) \exp\left(\frac{h_{0}}{2} \sum_{(i,j) \text{ pairs}}\hat{V}_{0ij}\right) \hat{\tilde{\psi}}_1 \\
&  \times \exp\left(\frac{h_{0}}{2} \sum_{(i,j) \text{ pairs}}\hat{V}_{0ij}\right) \exp\left(\frac{h_0}{2} \hat{H}_{\mathrm{Sun}}\right).\\
\end{aligned}
\label{eq:multsymp}
\end{equation}
$k_m$ is the maximum recursion level which, as in Section \ref{sec:imp}, may vary during a global timestep $h_0$.    $\v{W}_k$ is the vector of indices $i$ such that for $> k$, $\sum_{j \ne i} V_{lij}= 0$.  Here, $\sum_{(i,j) \text{ pairs}}$ indicates a sum over all pairs $(i,j)$ in some order which does not change the integrator due to eq. \eqref{eq:commute}.  As in eq. \eqref{eq:phi}, we define,
\begin{equation}
\begin{aligned}
\hat{\psi}_i =& \exp\left(h \left(\sum_{l = i}^\infty \sum_{0<j<k} \hat{V}_{ljk} + \sum_{j\in \v{W}_i} \hat{H}_{\mathrm{Kep},j} \right) \right)\text{  for $i \ne 0$, and}\\
\hat{\psi}_0 =& \exp\left(h \left(\hat{V} + \hat{H}_{\mathrm{Kep}} + \hat{H}_{\mathrm{Sun}} \right) \right),
\end{aligned}
\end{equation}
where $\hat{\tilde{\psi}}_i = \hat{\psi}_i + \mathcal O(h^2)$.  Similar to the situation in Section \ref{sec:imp}, the maximum recursion levels within $\hat{\tilde{\psi}}_0$ can only increase in our implementation.  We have implemented eq. \eqref{eq:multsymp}, {\bf which is the \texttt{SYMBA} map}, only for three-body systems, and call this implementation MTS again.  We do not implement this algorithm for systems with more particles as the estimates for the pairwise maximum recursion levels (Section \ref{sec:estmany}) have complications  and our focus here is not on symplectic algorithms.

\subsubsection{Estimating the maximum recursion level}
\label{sec:estmany}
For three bodies, we can apply the algorithm of Section \ref{sec:imp} to determine the maximum $h_i$ for the planetary pair.  Eq. \eqref{eq:Vk} simplifies to $V_{k12}$, and $H_{\mathrm{Kep}} =  H_{\mathrm{Kep,1}} + H_{\mathrm{Kep,2}}$.  We also make the substitutions $\v{q} \rightarrow \v{Q}_1 - \v{Q}_2$ and $\v{p} \rightarrow \v{v}_1 - \v{v}_2$, where $\v{v}_i = \v{P}_i/m_i$.  For systems with more bodies, it's unclear a pairwise maximum recursion level can be adapted within the global step while maintaining symplecticity and time-reversibility, so we do not attempt it.  The \texttt{SYMBA} code may have achieved this although we have not studied their implementation in detail. 

\subsection{Time-reversible integrators}
\label{sec:MRmany}

As in Section \ref{sec:tvar}, as a next novelty, we now develop a simpler, more flexible time-reversible version of the algorithm of Section \ref{sec:sympmany}.  First, given $\v{z}(0)$, identify the set of all pairs of particles $S_{k}$ at each timestep $h_k$ (based on some criteria like separation).  The set at the smallest timestep is $S_{k_{\mathrm{m}}}$.  Next, define maps, 
\begin{equation}
\hat{A}_k =  \exp\left(\frac{h_k}{2} \sum_{(i,j) \in S_k}\hat{\tilde{V}}_{ij} \right),  \qquad \mathrm{and} \qquad \hat{B}_k =  \exp\left({h_k  \sum_{i \in S_k^\prime} \hat{H}_{\mathrm{Kep},i}} \right).
\end{equation}
$\tilde{V}$ is defined from eq. \eqref{eq:Vk}.  The sum in $\hat{A}_k$ is over all pairs in $S_k$.  The same index can be present in different $S_k$.  $S^\prime_k$ is a 1D array of all the indices in $S_k$, but if any indices are present in $S_j$ for $j > k$, they have been removed.  In this way, $\hat{B}_k$ consists of Kepler operators for particles not treated at smaller timesteps $h_j$ (for $j > k$).  It also follows that, (for $j > k$) 
\begin{equation}
[\hat{B}_k,\hat{A}_j] = 0.  
\label{eq:commut}
\end{equation}
We now write down an evolution map,
\begin{equation}
\begin{aligned}
\hat{D}_{k_m} &= \left[ \hat{A}_{k_m} \hat{B}_{k_m} \hat{A}_{k_m}      \right]^M,\\
\hat{D}_i &= \left[ \hat{A}_{i}  \hat{D}_{i+1} \hat{B}_i \hat{A}_{i}      \right]^M\text{  for $0 < i < k_m$, and}\\
\hat{D}_0 &= \exp\left(\frac{h_0}{2} \hat{H}_\mathrm{Sun}\right) \hat{A}_0 \hat{D}_1 \hat{B}_0 \hat{A}_0 \exp\left(\frac{h_0}{2} \hat{H}_\mathrm{Sun}\right).
\end{aligned}
\label{eq:Cmap}
\end{equation}
$\hat{D}_0$ generalizes map \eqref{eq:Ckep}.  Due to property \eqref{eq:commut}, $\hat{D}_0$ is time-symmetric.  We can now write generalizations of the reversible algorithms MTR and AG, Listings \ref{lst:adapglobal} and \ref{lst:adapth0}, that work with more than two bodies.  For MTR, \texttt{mapC} is given by eq. \eqref{eq:Cmap}.  Before \eqref{eq:Cmap} is applied, we construct a timestep level matrix \texttt{iN0}, indicating each pair's timestep level.  \texttt{iN0} has size $n(n-1)/2$, with $n$ the number of planets.  For problems with large $n$, not considered in this paper, manipulating an array of this size may become expensive, and a more efficient computational approach could be devised.  We apply \eqref{eq:Cmap} and timestep levels are computed at each substep for each planet pair.  The maximum level for each pair is recorded as \texttt{iN}, substituting Line \texttt{07} of Listing \ref{lst:adapglobal}.   If any level has increased compared to \texttt{iN0}, the step is redone.  If close encounters occur frequently, one can expect the compute effort to, at most, double compared to an algorithm that does not redo steps.  This is because redoing steps more than once is a rare case in our tests.  After all maps are applied, \texttt{iNt} is calculated at the final coordinates and momenta, to be used at the beginning of the next global timestep.

To generate AG, in Line \texttt{07} of Listing \ref{lst:adapth0}, $j$ must be the minimum over all planetary pairs.  The leapfrog method, \texttt{mapphi0}, is replaced by the Wisdom--Holman map, eq. \eqref{eq:psi2}.  Adapting the global timestep alone with AG will only be efficient for problems in which one pair of planets has close encounters at a time (more precisely, we consider it a close encounter if the pair timestep is smaller than $h_0$).  For problems in which different pairs of planets have simultaneous close encounters, for efficiency, we should use MTR.  An even more efficient approach would be to simultaneously adapt the pairwise steps with MTR and adapt the global step with AG, as needed.  For many problems which are reasonably stable, a global timestep can be kept fixed, but if the system settles into a state with slower or faster timescales, adapting $h_0$ may make sense.

\subsection{Comparing the error analysis for two algorithmic approaches for close encounters}
\subsubsection{Hybrid symplectic integrators}
In this work, we have focused on decreasing pairwise timesteps during planetary close encounters.  An alternate symplectic strategy was presented by \cite{C99}, which we describe here, and we study its error.  \cite{C99} describes a hybrid symplectic integrator to resolve close encounters.  The idea is to use a cheap map $\hat{M}_1$ when there is no close encounter and switch to a more expensive map $\hat{M}_2$ during close encounters.  $\hat{M}_1$ can be written, 
\begin{equation}\label{eq:WH:maps}
	\hat{M}_1 =
	\Exp{\frac{h}{2}\sub{\op{B}}{1}}
	\Exp{h\sub{\op{A}}{1}}
	\Exp{\frac{h}{2}\sub{\op{B}}{1}},
\end{equation}
where $\sub{\op{A}}{1} = \hat{H}_{\mathrm{Kep}}$ and $\sub{\op{B}}{1} = \hat{H}_{\mathrm{Sun}} + \hat{V}$.  If $\epsilon = \max(m_i/m_0)$, it is easy to show $B_1/A_1 = \mathcal O(\epsilon)$.  For $\hat{M}_1$, we find,
\begin{equation}
	\label{eq:WH:Herr:BAB}
	\sub{H}{err}^{M_1} =
	+\frac{h^2}{12}\{\{\sub{B}{1},\sub{A}{1}\},\sub{A}{1} + \frac{1}{2} B_1\}
	+ \mathcal O(h^4),
\end{equation}
and it follows $\sub{H}{err}^{M_1} = \mathcal O(\epsilon h^2)$.  During a very close encounter, this error scaling becomes only $\mathcal O(h^2)$ so a hybrid method switches to map $\hat{M}_2$, where, 
\begin{equation}\label{eq:WH:maps}
	\hat{M}_2 =
	\Exp{\frac{h}{2}\sub{\op{B}}{2}}
	\Exp{h\sub{\op{A}}{2}}
	\Exp{\frac{h}{2}\sub{\op{B}}{2}},
\end{equation}
and $\sub{\op{A}}{2} = \hat{H}_{\mathrm{Kep}} + \sum_{(i,j) \in W}\hat{\tilde{V}}_{ij}$ and $\sub{\op{B}}{2} = \hat{H}_{\mathrm{Sun}} + \sum_{(i,j) \not\in W} \hat{\tilde{V}}_{ij}$.  $W$ is the set of pairs in a close encounter.  Switching to $\hat{M}_2$ allows hybrid symplectic integrators to maintain excellent error behavior.  A disadvantage is that $A_2$ needs to be solved via expensive methods like Bulirsch--Stoer.  
\subsubsection{Symplectic multiple timestep algorithms}
It is straightforward to write the error Hamiltonian of $\hat{\tilde{\psi}}_0$; this was also derived in eq. (35) of DLL98:
\begin{equation}
\begin{aligned}
H_{\mathrm{err}}^{\mathrm{DLL98}} =& \frac{h_0^2}{12}\left\{ \{H_{\mathrm{Sun}},H_{\mathrm{Kep}} \}, H_{\mathrm{Kep}} + \frac{1}{2} H_{\mathrm{Sun}}   \right\} \\
+& \sum_{i = 0}^\infty \frac{h_i^2}{12}\left\{ \{V_i, H_{\mathrm{Kep}} \}, H_{\mathrm{Kep}} + \frac{1}{2} V_i + \sum_{j = i+1}^\infty V_j   \right\} + \mathcal O(h_i^4)
\end{aligned}
\label{eq:Herrsymba}
\end{equation}
In the absence of close encounters, $H_{\mathrm{err}}^{\mathrm{DLL98}} = \mathcal O(\epsilon h_0^2)$.  We now further simplify eq. \eqref{eq:Herrsymba}, as we did in eq. \eqref{eq:Herrsimp}.  The force experienced by $i$ due to $j$ is found by, 
$\v{F}_{kij} = -{\partial {V}_{kij}}/ {\partial \v{Q_{i}}} = \tilde{\v{F}}_{kij}  - \tilde{\v{F}}_{(k-1)ij}$.  If $Q_{ij} \approx r_{k+1}$,  $\v{F}_{kij} = -\v{Q}_{ij} G m_i m_j/Q_{ij}^3 $ and $\v{F}_{(k-1)ij} = \v{F}_{(k+1)ij} = \v{0}$.  During a close $(i,j)$ encounter, the dominant contribution to the error Hamiltonian is,
\begin{equation}
H_{\mathrm{err}}^{(i,j)} = \frac{h_k^2}{24}\left\{ \{V_{kij}, H_{\mathrm{Kep}} \}, V_{kij}   \right\} + \mathcal O(h_k^4).
\label{eq:Herrsymbap}
\end{equation}
Note the difference with eq. \eqref{eq:Herrsimp}.  We can see that $H_{\mathrm{err}}^{(i,j)} = \mathcal O(h_k^2 Q_{ij}^{-4})$.  By choosing $R = \sqrt{M}$, $H_{\mathrm{err}}^{(i,j)}  = \mathcal O(h_0^2)$, independent of the strength of the $(i,j)$ close encounter.  With $M=3$ and $R=2$, $H_{\mathrm{err}}^{(i,j)} = \mathcal O(h_0^2 Q_{ij}^{-1})$, adequate if the close encounter is not extreme. 
\subsubsection{Discussion}
The hybrid and multiple timestep integrators are two different algorithmic approaches to ensuring the error of the Wisdom--Holman method remains small during planetary close encounters.  The approach of hybrid methods is more straightforward, but requires use of expensive conventional integrators like Bulirsch--Stoer during close encounters.  Conceptually, it is unappealing to use a conventional integrator within a symplectic integrator.  Multiple timestep algorithms can rely entirely on Kepler solvers and need not invoke conventional integrators.  Conceptually, it is also unappealing to take the Wisdom--Holman method, which is designed under the assumption of no close encounters occurring, and to use brute force and decrease its timestep, to make it able to handle close encounters.  Clearly, there is no perfect solution to the problem of planetary close encounters in $N$-body symplectic simulations.

\subsection{Numerical comparisons among algorithms}
We now present numerical experiments that demonstrate the performance of the reversible methods for systems with more than one planet (with a dominant mass).  

\subsubsection{The restricted three-body problem}
\label{sec:3bod}
\begin{figure}
	\includegraphics[width=90mm]{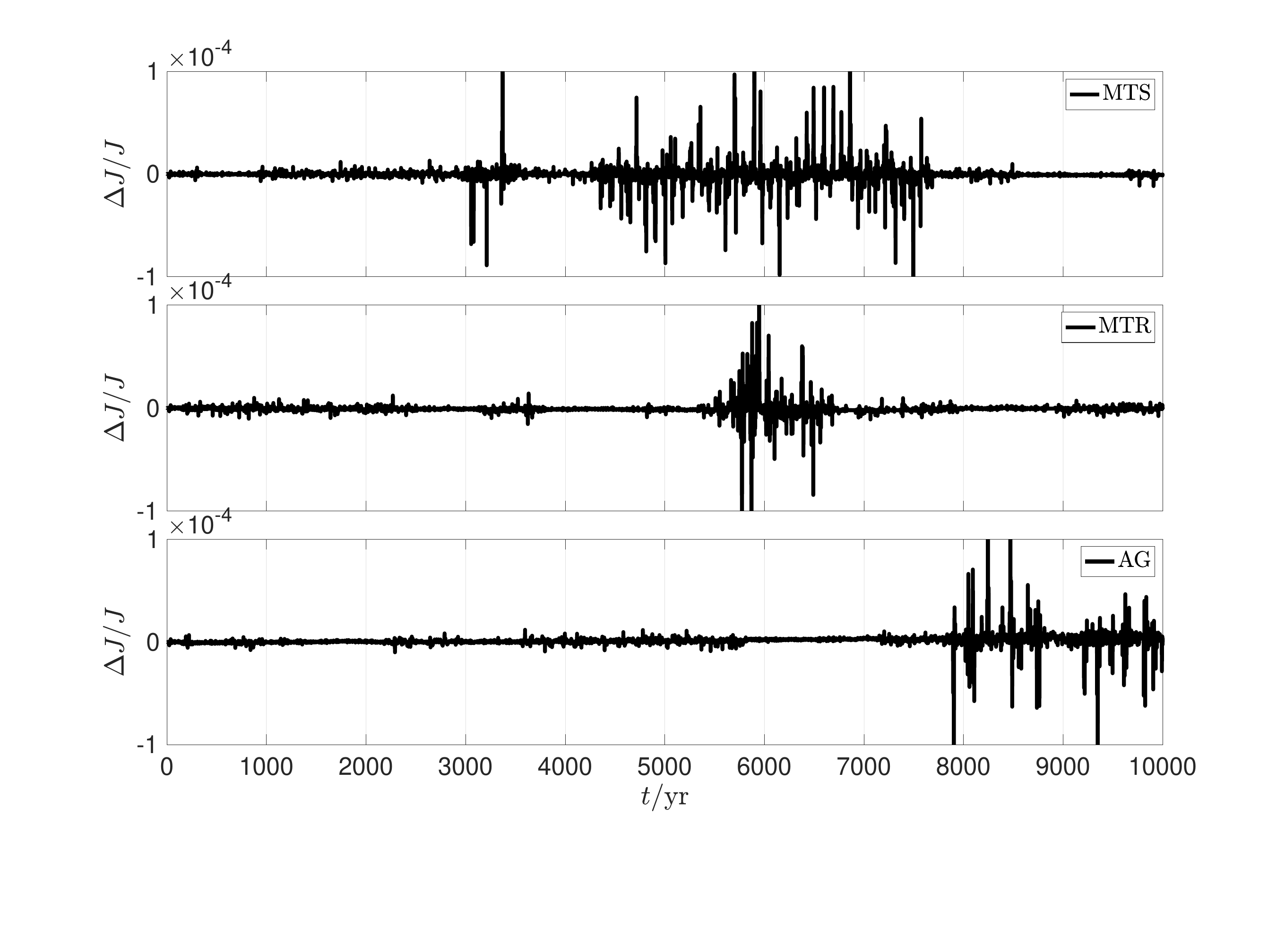}
	\caption{
		A comparison of the error performance of the symplectic and reversible methods for a planar restricted three-body problem, in which the test particle can visit primary and secondary and has many close encounters with secondary.  We record the error in the Jacobi constant over $843$ secondary periods.  The errors of all methods are comparable and the fastest method is the reversible AG method.
	\label{fig:3bod}
  	}
\end{figure}

As one of the simplest chaotic systems with more than one planet, we consider the planar, restricted three-body problem, in which there is one conserved quantity, the Jacobi integral.  Our test problem is taken from \cite{Wisdom2017}, and consists of the ``Sun,'' ``Jupiter,'' on a circular orbit, and a test particle with initial conditions such that it can visit the other two masses.  The test particle has multiple close encounters with Jupiter.  For example, in one integration, there were $34$ close encounters in the first $1000$ years.  Here, a close encounter is defined by the (only) timestep level jumping from $1$ to a higher level, and back down to $1$.  A timestep level of $1$ is the global timestep $h_0$.  As there is only one planetary pair in this problem, we can also apply AG to it.  We set $h_0 = 8$ days, run for $10000$ yrs, set $r_1 = 5 r_{\mathrm{Hill}}$, with $r_{\mathrm{Hill}}$ the Hill radius of Jupiter, and $M=4$ and $R=2$ for all methods.  Output is generated at one year intervals (Jupiter's period is $11.86$ yrs).  We compare the error in the Jacobi integral over time for MTS, MTR, and AG in Fig. \ref{fig:3bod}.  
The errors of all methods do not drift significantly and have similar median errors, respectively: $-2.3 \times 10^{-7}$, $-9.5 \times 10^{-7}$, and $1 \times 10^{-6}$.  The runtimes are $256$, $280$ and $145$ s, respectively: the fastest method is again AG with compute time about $57\%$ that of MTS.  Note we are deliberately avoiding comparing compute times with codes like \texttt{MERCURIUS} or \texttt{SYMBA}, since our implementations are not written in fast languages.

\subsubsection{Violent Solar System}
\label{sec:viol}
Next, we will test a problem demonstrating the remarkable difference a few irreversible timesteps can have on the accuracy of the evolution of a dynamical system.  We consider a violent outer Solar System, introduced by DLL98.  We reuse the following parameters: the system configuration of Sun plus outer giant planets, with planetary masses scaled up by a factor $50$; $h_0 = 0.03$ yrs for MTR, and a runtime of $3000$ yrs.  To match the DLL98 experiment as closely as possible, we use position dependent shells with $r_1 = 1.52$ au, about $3$ initial Saturn Hill radii.  $r_1$ is used for all pairs.  We set $R=2$ as in DLL98.  However, in contrast to DLL98, we use $M=4$ rather than their $M=3$.  We were unable to achieve good accuracy with their choice for both MTR and AG.  This can have many explanations; in particular as we note below, our dynamical evolution of this chaotic system differs from that of DLL98; note our initial conditions are different and the Lyapunov time in this problem can be short; e.g., $40$ yrs \citep{H16}. We have also discarded the possibility that higher accuracy can be attributed to the maximum recursion level algorithm of Section \ref{sec:imp}.  To check this, we implemented a higher accuracy version of MTR in which we not only collect the separations between substeps, but also the minimum separation within substeps, to determine timestep levels.  There was no improvement in any result.  We also tested a Kepler problem and found that checking minimum separations within substeps in no case altered recursion levels for the symplectic method.  So checking separations within timesteps leave both the symplectic and reversible results unaltered in our tests.  From previous experience, we do not expect that \texttt{SYMBA}'s smooth switching function, which contrasts to our discrete switching function, is responsible for differences in accuracy.  

\begin{figure}
	\includegraphics[width=90mm]{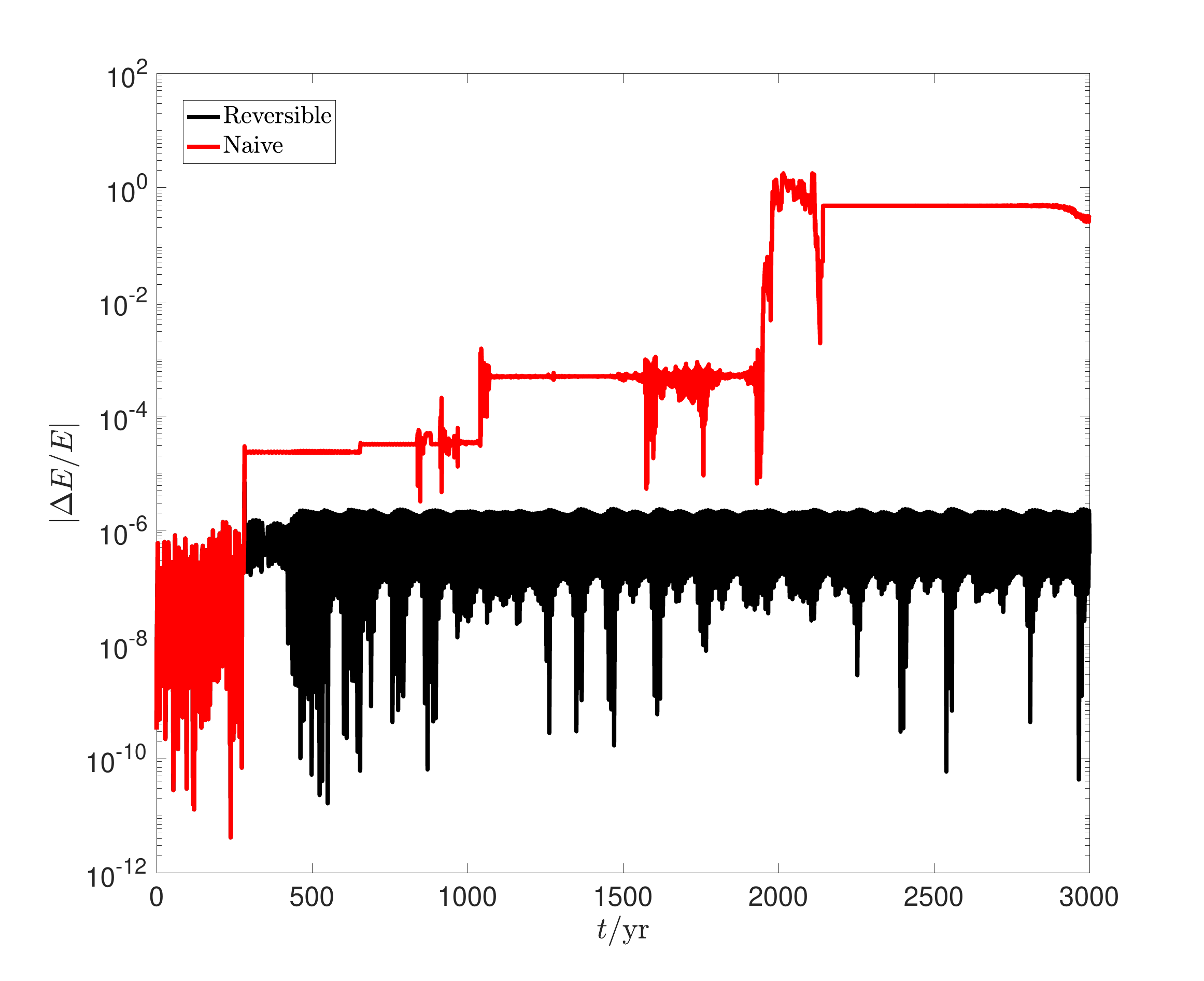}
	\caption{
	Error over time of a time integration of a violent outer Solar System, consisting of the Sun and outer giant planets, whose masses have been scaled up by a factor $50$.  We compare the MTR method (``Reversible''), in which $8$ steps have been repeated (a fraction $8 \times 10^{-5}$), to the same method without any repetition (``Naive'').  While ``Naive'' reaches an error of order unity, MTR maintains an error to about $2$ parts in a million from redoing the $8$ steps.
	\label{fig:violentE}
  	}
\end{figure}

In Fig. \ref{fig:violentE}, we compare the performance of MTR (labeled ``Reversible'')  and a naive, non-reversible method (labeled ``Naive''), which is simply MTR with no repeated steps.
MTR only redoes $8$ steps in the entire simulation, a fraction of $8 \times 10^{-5}$, and is almost like the naive method in the sense that it rarely repeats steps.  Although the number of close encounters is small for this tests, we know from our long-term extreme eccentricity Kepler orbit tests of MTR (Section \ref{sec:e999}) that good accuracy is maintained even with more frequent encounters.  The MTR error is similar to that obtained with \texttt{SYMBA} as reported by Fig. 6 of DLL98.  MTR's error oscillates mostly between $1$ and $-2 \times 10^{-6}$, while theirs oscillates between about $2$ and $-4 \times 10^{-6}$.  For the naive method, the error blows up to order unity by the end of the integration.  The naive and reversible curves are identical until $t = 281$ yrs.

\begin{figure}
	\includegraphics[width=90mm]{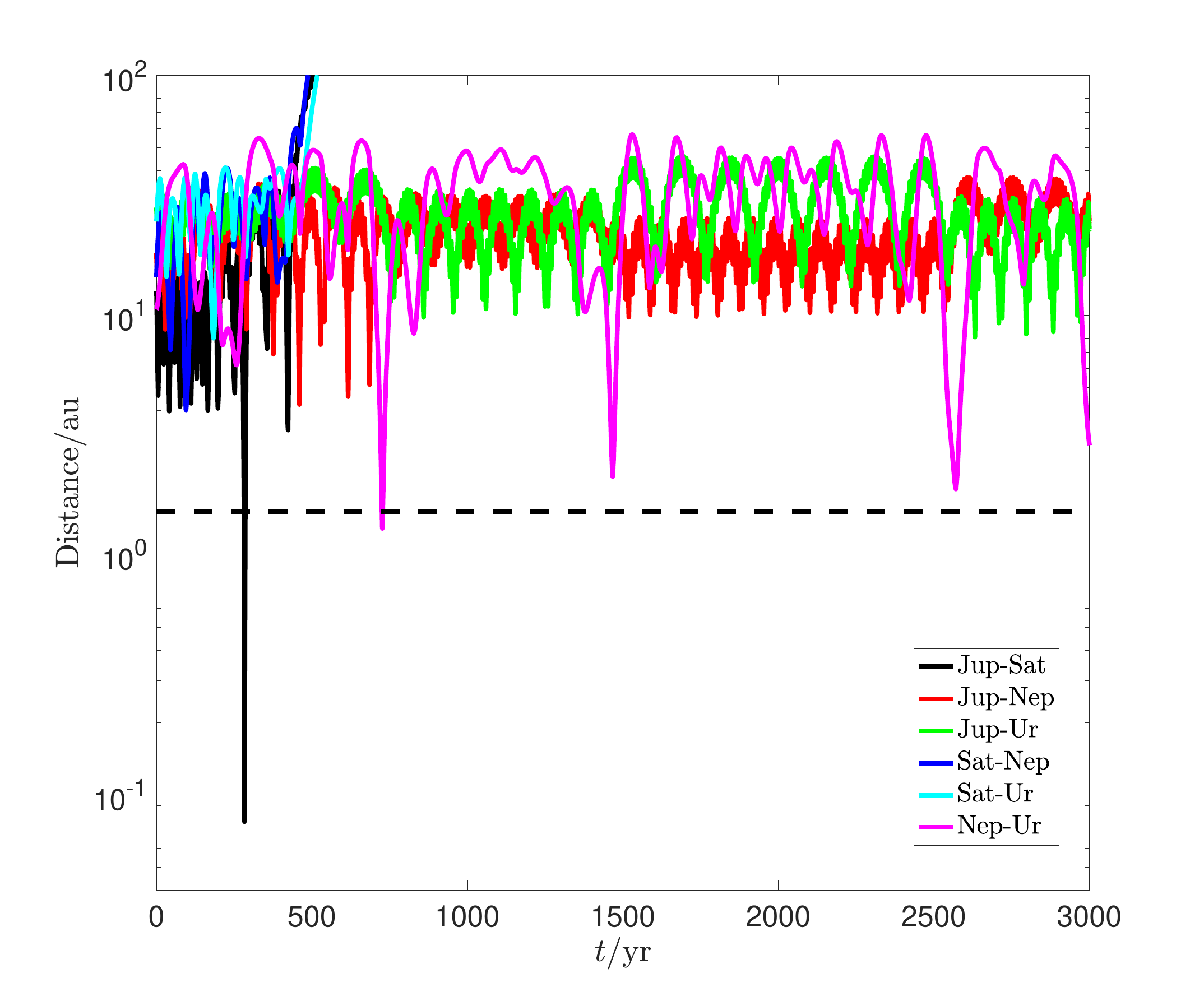}
	
	\caption{
	Pairwise distances over time for the system of Fig. \ref{fig:violentE}, obtained with the reversible method.  Saturn's initial Hill radius is indicated by a dashed line.  There is a close encounter between Jupiter and Saturn at $282$ yrs.  Then there are three additional close encounters between Uranus and Neptune.  Saturn is ejected at $\approx 430$ yrs.
	\label{fig:violentQ}
  	}
\end{figure}

To more carefully understand the dynamical behavior, Fig. \ref{fig:violentQ} plots the six planetary pairwise distances as a function of time for MTR.
Saturn's initial Hill radius is indicated with a dashed line.  The Jupiter--Saturn separation reaches a minimum of $< 0.077$ au at $282$ yrs (because the separation is measured at output, $0.077$ au is an upper bound).  This is $160$ Jupiter radii.  MTR attempts a timestep level of $6$, corresponding to $h = 15.9$ min.  Saturn is ejected at $\approx 430$ yrs, which contrasts to the result in DLL98, where it is ejected at $1950$ yrs.  Uranus and Neptune have a series of three close encounters in which the timestep level jumps up to $2$.  Besides these four events, no other timestep levels ever jump past $1$.  In contrast to the result of DLL98, Uranus is never ejected.  As discussed above, local Lyapunov times for this system can be small, and our initial conditions likely differ from those of DLL98, so the different dynamical evolution is not surprising.

\subsubsection{Planetary system with hierarchical binaries}
\label{sec:5bod}
As a final test, we wish to simulate a complicated system in which different pairs of particles are at different varying timestep levels simultaneously.  We use MTR to study this system.  AG alone would not give us the flexibility to adapt all pairwise steps.  To find such a system, we build on the binary planet problem introduced by DLL98, Section 6.1.  We have a Solar mass star with one binary planet placed at $1$ au and another binary planet at $3$ au, for a total of five bodies.  The first binary has $a = 0.0125$ au and $e = 0.6$, while the second binary has $a = 0.013$ au and $e = 0.2$.  All planets have mass $10^{-3} M_{\odot} $.  As in DLL98, Section 6.1, we set $h_0 = 0.01$ yr and integrate for $100$ yrs, so that the total number of periods in the system for the different hierarchical binaries are $19$, $100$, $3017$, and $3200$.  Although we could construct velocity-dependent timestep levels for this problem, in contrast to the case of \texttt{SYMBA}, we find instead that timestep levels based on the relative free-fall times $t_{\mathrm{ff}}$ work well.  For this system, we prefer this to timestep levels based on Hill radii as proposed by DLL98, because the Hill radius is less well motivated.  For pair of planets $(i,j)$, $t_{\mathrm{ff}} = [Q_{ij}^3/(G (m_i + m_j)) ]^{1/2}$.  

To get the timestep levels, use shells of a nondimensional function $g = t_{\mathrm{ff}}/h_0$, and let $g_i/g_{i+1} = R$, in analogy to the radius ratios explored in Section \ref{sec:MT}.  We set $g_1 = 30$, which was determined numerically to work well.  We set $R = 2$ and $M = 3$.  

\begin{figure}
	\includegraphics[width=90mm]{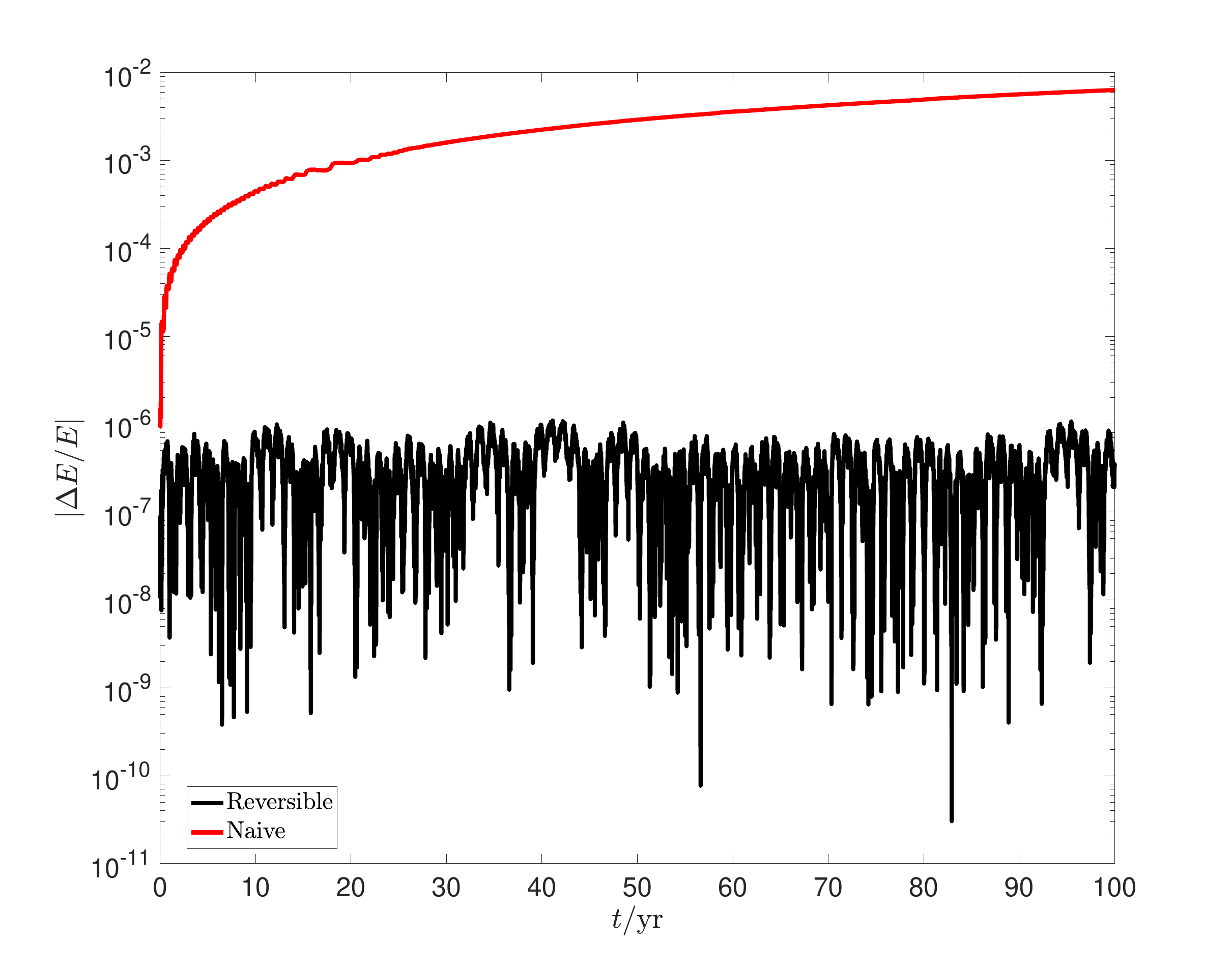}
	
	\caption{
	Error over time for an integration of a hierarchical five-body problem, with the two methods of Fig. \ref{fig:violentE}.  The system consists of a Solar mass star, a binary planet at $1$ au, and another binary planet at $3$ au, with both pairs in eccentric orbits.  The pairwise timestep levels vary substantially, as shown in Fig. \ref{fig:bin2}.  The integration time is $100$ yrs, over which the binary planets take about $3200$ orbits.  The reversible method conserves energy to better than a part in a million, while the naive method's error is worse than $0.1\%$.
	\label{fig:bin2E}
  	}
\end{figure}

\begin{figure}
	\includegraphics[width=90mm]{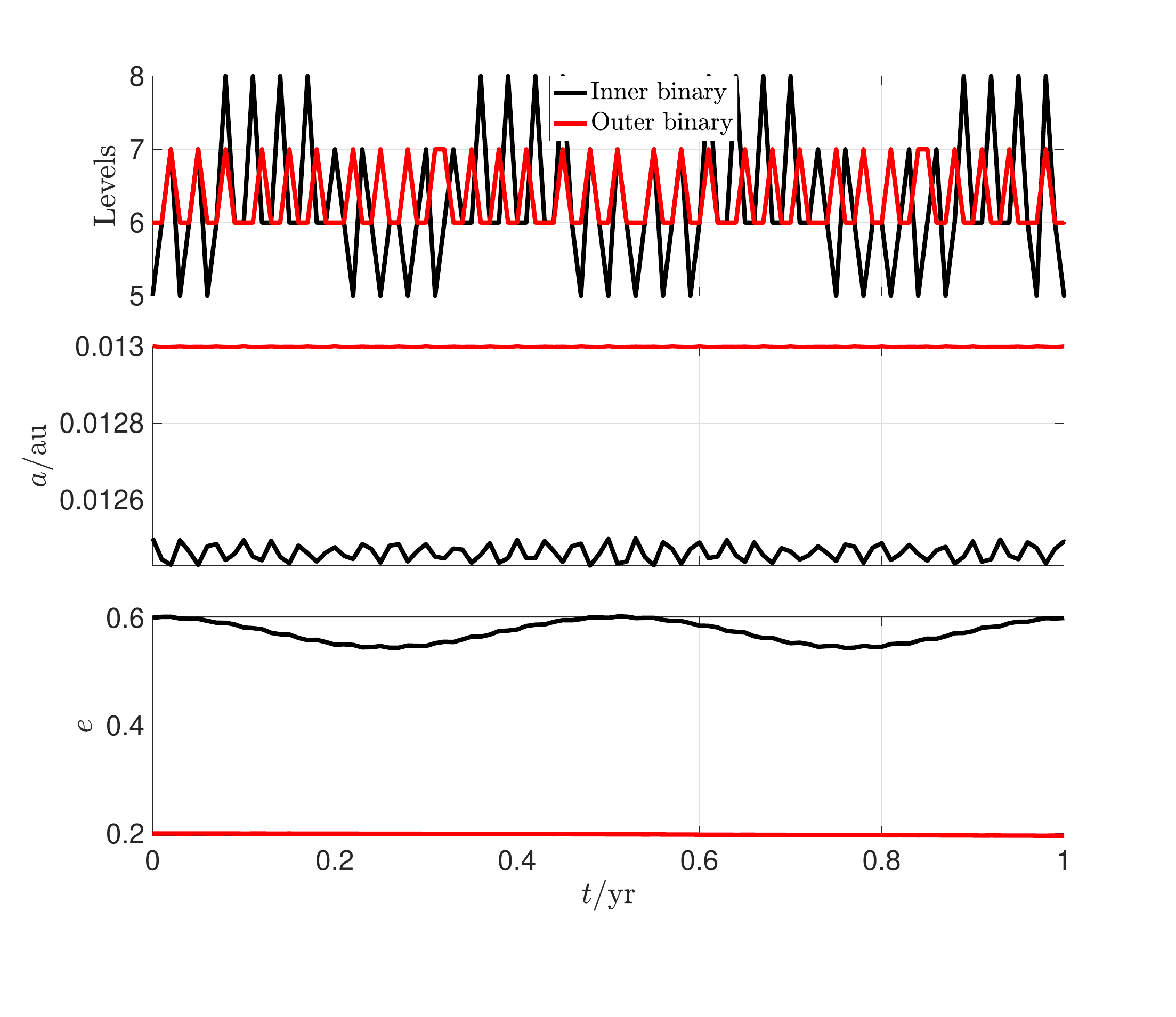}
	\caption{
		Attempted timestep levels, semi-major axes, and eccentricities for the pair of binary planets of the system of Fig. \ref{fig:bin2E}.  The results of MTR are used to make this pot which shows $1\%$ of the time of Fig. \ref{fig:bin2E}.  The two pairs have timestep levels varying independently, which MTR handles well.
	\label{fig:bin2}
  	}
\end{figure}

Fig. \ref{fig:bin2E} plots the energy error over time again for a reversible and a naive method.  For the naive method, the error grows linearly in time by about four orders of magnitude.  The error is well preserved for the reversible method.  
Fig. \ref{fig:bin2} displays $a$ and $e$ for the two planetary binaries in one year using the reversible method.  
The elements do not drift significantly.  We also show the timestep levels of the two binaries.  Between global timesteps, the inner binary's attempted step level varies by $\pm 3$, between $5$ and $8$.  The inner binary actually uses timestep levels between $6$ and $8$ that change by at most $\pm 2$.  For the outer binary, the variation is $\pm 1$, between $6$ and $7$, both for attempted and used timestep levels.  All other planetary pairs remain at level $1$ ($= h_0$).  $67\%$ of steps were redone, while a fraction $10^{-4}$ steps were redone twice.  No steps were redone more than twice.  

Thus, we showed that the reversible method can handle timestep level changes by more than $\pm 1$ in a global timestep, and pairwise timestep levels can vary independently, all without incurring secular energy error drifts.  Finally, we comment that the reversible method we used for this system might be made more efficient still by combining Listings \ref{lst:adapglobal} and \ref{lst:adapth0p}, to adapt the pairwise steps locally in time; exploration of this is left for future work.
\section{Conclusions}
\label{sec:conc}

We presented two time-reversible algorithms, AG (Adaptive Global) and MTR (Multiple Timesteps Reversible).  MTR can be understood as a Wisdom--Holman method \citep{WH91} (WH) in which the planetary pairs can have their own independently varying timestep, increasing the WH efficiency.  When adapting the timesteps, a global timestep is simply redone if it is determined to not be reversible; the repeated timesteps are often a small fraction of all steps.  AG is WH in which its global timestep can be adapted at block-synchronized times, without losing any of its beneficial long-term error properties.  Indeed, AG can simply replace WH in most circumstances to increase its efficiency by making it more flexible and adaptable.  These methods combined can be considered a simple, flexible, and adaptive version of the multiple timescale algorithm \texttt{SYMBA}, presented in DLL98.  We implemented a symplectic algorithm based on \texttt{SYMBA} which we call MTS (Multiple Timestep Symplectic).  We tested the symplectic and reversible algorithms on various challenging planetary $N$-body problems exhibiting close encounters.  The accuracy of the reversible algorithms was as good as MTS and \texttt{SYMBA} (as seen in the figures of DLL98), and AG was up to $2.6$ times faster than MTS.  In contrast to \texttt{SYMBA}, the reversible methods presented here require no switching functions and are able to adapt timestep levels based on not only the pairwise separations between bodies.

To construct the reversible methods, we generalized the results of \cite{HD2023}, who show how to switch reversibly between two timesteps.  Our new algorithms instead can switch between an arbitrary number of maps, each with different pairwise planetary timesteps or global timesteps.  The methods simply change timesteps when needed and then we check if the global timestep was reversible.  If not (often in a small fraction of cases), the global timestep is redone, resulting in an ``almost'' reversible method \citep{HD2023}.  For maximum efficiency, MTR and Listing \ref{lst:adapth0p} (not explored in this work) can be combined, so that steps are adapted locally in time, and we hope to present this implementation in future work.  The codes studied in this paper are available for reference on Github\footref{Rstep}, and are written in an interpreted language.  We leave a user friendly implemented in a compiled language for future work.

Our tests were limited to challenging systems with at most five bodies, including hierarchical binaries and a violent outer Solar System, but we expect the algorithms to work well in large particle systems, which can be studied in a language like C and is, again, left for future work.  We remark that we have assumed conservative Hamiltonian systems when deriving our methods, but we can apply the methods here to conservative systems with nonconservative perturbations.

Taking a broader view, we see that traditional symplectic methods can be made simpler and more flexible simply by translating them to a reversible framework.  While we have limited ourselves to planetary dynamics problems, there is no reason not to believe we can similarly transform other symplectic methods in other domains of astronomy.

We also remark on a significant finding in the course of this work, even if it is not the main result.  Adapting the global timestep of any time-symmetric (including symplectic) algorithm via the simple prescription here, which builds on \cite{HD2023}, can be immediately applied to many codes.  However, it is curious that we found that, while the timestep can be arbitrarily decreased, it can only be increased at block synchronized times, or similar prescriptions.  This should not be viewed as a limitation of AG because it should not really affect performance, but rather a feature of algorithms attempting to discretely vary the stepsize.  Without taking care of this, the long-term error performance will be destroyed.  This finding contrasts with the results of \cite{HMM95}, who vary timesteps reversibly and continuously but find no similar requirement.  We note the \cite{HMM95} algorithm is implicit, rendering it more expensive to use.  

\section{Acknowledgements}
\label{sec:ack}
We thank Man Hoi Lee for help with \texttt{SYMBA}.  We thank Tiger Lu for a careful reading.  We thank the reviewer, John Chambers, for thoughtful comments.


\section{Data Availability}
Example implementations of the algorithms used in the experiments in this paper are available at \texttt{https://github.com/dmhernan/Reversible-Stepping}.  The data underlying this article will be shared on reasonable request to the author.

\bibliographystyle{mnras}
\bibliography{paper}

\end{document}